# A BeiDou Signal Acquisition Approach Using Variable Length Data Accumulation based on Signal Delay and Multiplication

**Menghuan Yang [1,2], Hong Wu [1,2,\*], Qiqi Wang [3], Yingxin Zhao [1,2], and Zhiyang Liu [1,2]**

1. College of Electronic Information and Optical Engineering, Nankai University, Tianjin 300350, China; 1120150087@mail.nankai.edu.cn (M.Y.); zhaoyx@nankai.edu.cn (Y.Z.); liuzhiyang@nankai.edu.cn (Z.L.)
2. Tianjin Key Laboratory of Optoelectronic Sensor and Sensing Network Technology, Tianjin 300350, China
3. College of Electronic Information and Automation, Tianjin University of Science & Technology, Tianjin 300222, China; wangqiqi@tust.edu.cn
\* Correspondence: wuhong@nankai.edu.cn

**Abstract:** The secondary modulation with the Neumann–Hoffman code increases the possibility of bit sign transition. Unlike other GNSS signals, there is no pilot component for synchronization in BeiDou B1/B3 signals, which increases the complexity in acquisition. A previous study has shown that the delay and multiplication (DAM) method is able to eliminate the bit sign transition problem, but it only applies to pretty strong signals. In this paper, a DAM-based BeiDou signal acquisition approach, called variable length data accumulation (VLDA), is proposed to acquire weak satellite signals. Firstly, the performance of DAM method versus the different delays is analyzed. The DAM operation not only eliminates bit sign transition, but it also increases noise power. Secondly, long-term signal is periodically accumulated to improve signal intensity in order to acquire weak signals. While considering the Doppler frequency shift of ranging codes, the signal length must be compensated before accumulating long-term signal. Finally, the fast-Fourier-transform based parallel code phase algorithm are used for acquisition. The simulation results indicate that the proposed VLDA method has better acquisition sensitivity than traditional non-coherent integration method under the same calculation amount. The VLDA method only requires approximately 27.5% of calculations to achieve the same acquisition sensitivity (35 dBHz). What is more, the actual experimental results verify the feasibility of the VLDA method. It can be concluded that the proposed approach is an effective and feasible method for solving the bit sign transition problem.

## 1. Introduction

The BeiDou Navigation Satellite System (BDS) is a Global Navigation Satellite System (GNSS) constructed and operated by China. The ambitious BDS constellation consists of Geostationary Earth Orbit (GEO) satellites, Inclined Geosynchronous Orbit (IGSO) satellites, and Medium Earth Orbit (MEO) Satellites [1]. So far, China has launched 53 satellites [2], and more than 40 of which are currently transmitting navigation signals [3]. BDS will play an important role in high-accuracy positioning, navigation, and timing service.

Acquisition is the first step of the BeiDou receiver baseband processing. While considering the huge path loss, it is not an easy task to receive satellite signals on the earth. The minimum received power levels on the ground of the BeiDou signals are specified to be -163dBW [4–5]. The power will be less if the satellite signals are blocked. Under this circumstance that the satellite signal is submerged in the background noise, it requires long term coherent integration to implement signal acquisition. However, the bit sign transition will cause the correlation peaks to be neutralized during the long-term coherent integration. The higher symbol rate of the satellite signals, the more frequently bit sign will occur.

In BeiDou B1/B3 signals, the D1 navigation message is secondary modulated with the Neumann–Hoffman (NH) code [4–5]. The secondary code is also used in other GNSS signals, including the modernized GPS, Galileo, and GLONASS. The NH codes extend the period of the spreading code, which reduces the interval of the spectrum line and suppresses the narrowband interference. On the other hand, NH encoding increases the symbol rate from 50bps to 1kbps, which increases the acquisition difficulty. Table 1 summarizes the signal structures of the typical GNSS signals that are modulated with secondary code. All information comes from the official Interface Control Documentary (ICD) [4–8]. Unlike other GNSS signals that are modulated with NH code, the BeiDou B1/B3 signal does not have a pilot component that is used for synchronization. It makes the acquisition of BeiDou B1/B3 signals more difficult than that of other GNSS signals.

**Table 1.** Structures of the typical Global Navigation Satellite System (GNSS) signals modulated with secondary code.

| GNSS System | BDS* | | | GPS | | GLONASS | | Galileo | |
|---|---|---|---|---|---|---|---|---|---|
| Service signal | B1I | B2a | | B3I | L5C | | L3OC | | E1 OS |
| Signal component | data | data | pilot | data | data | pilot | data | pilot | data | pilot |
| ranging code length | 2046 | 10230 | 10230 | 10230 | 10230 | 10230 | 10230 | 10230 | 4092 | 4092 |
| ranging code rate, Mcps | 2.046 | 10.23 | 10.23 | 10.23 | 10.23 | 10.23 | 10.23 | 10.23 | 1.023 | 1.023 |
| Secondary code length | 20 | 5 | 100 | 20 | 10 | 20 | 5 | 10 | — | 25 |
| Secondary code rate, Kcps | 1 | 1 | 1 | 1 | 1 | 1 | 1 | 1 | — | 250 |
| Symbol rate, kbps | 1 | 1 | 1 | 1 | 1 | 1 | 1 | 1 | — | 250 |
| Data rate, bps | 50 | 200 | — | 50 | 50 | — | 100 | — | 250 | — |

* The D1 navigation message broadcast by MEO/IGSO satellites is modulated with secondary code, while the D2 navigation message broadcast by GEO satellites is NOT modulated with secondary code.

Overcoming the effect of bit sign transition has become a critical issue in BeiDou signal acquisition [9–15]. A previous study has evaluated eight classical acquisition methods [16]. The FFT-based parallel search algorithm [17] is considered as the standard method. While considering the high efficiency of FFT, the FFT-based parallel search algorithm is widely accepted in the currently used acquisition methods [18]. Besides, non-coherent (NCH) integration and differentially coherent (DFC) integration are useful methods for weak signal acquisition [19–22]. The sensitivity of acquisition will be improved by extending the coherent integration length. Double block zeros padding (DBZP) is a useful method for suppressing bit sign transition [13, 14, 23, 24]. It was first introduced in GPS P(Y) code acquisition and has been well tested in acquisition of GPS and Galileo system [24]. DBZP method has great potential to acquire pretty weak BeiDou signal as weak as 22 dBHz [14]. However, the DBZP method is generally considered to have low efficiency because half of the signal power is wasted, which increases the computation complexity [25]. Furthermore, many efforts have been made in block-level correlation to made improvements in the DBZP method [24–27].

Previous research has shown that the delay and multiplication (DAM) method can eliminate bit sign transition [16, 28, 29]. By multiplying the intermediate frequency (IF) signal with a delayed version of itself, the carrier and bit sign transition are all removed. It is considered to be the fastest acquisition approach when the signal is strong enough [16]. The weakness of DAM is that the noise increases sharply when two signals with noise are multiplied. It is believed that the DAM method cannot find weak signals [29]. The DAM method has not been thoroughly studied and practically applied due to the fatal flaw mentioned above.

In this paper, we have developed the potential of the DAM method to acquire weak signals. We propose a BeiDou signal acquisition approach while using variable length signal accumulation (VLDA) based on DAM. The proposed method consists of three steps. Firstly, the BeiDou signals are delayed and multiplied to eliminate the carrier and bit sign transition. Secondly, the long-term signals are periodically accumulated to improve signal intensity. Finally, the FFT-based parallel code phase search algorithm is used for acquisition.

The highlights of this paper are summarized, as follows.

- The noise performance of the signal after DAM operation is analysed. The delay in DAM operation is optimized to minimize the noise power, maximize signal power, and optimize the correlation performance of ranging codes.
- The VLDA method is proposed to improve the signal strength after DAM operation.
- The simulation results show that the proposed VLDA method has better acquisition sensitivity than traditional NCH method under the same calculation amount. The VLDA method requires only about 27.5% of calculations to achieve the same acquisition sensitivity (35dB-Hz).

The rest of this paper is organized, as follows. Section 2 analyses the principle of DAM and the proposed VLDA algorithm. Section 3 provides the simulation experiment results. Section 4 discusses the advantages and disadvantages of the proposed VLDA method. Finally, Section 5 presents the conclusion.

## 2. Principles and Methods

### 2.1. The Principle of DAM Method

The main purpose of the DAM method is to eliminate the bit sign transition and the carrier. This method is very interesting from a theoretical point of view. However, it has not been widely used, because it only applies to a pretty strong signal. The principle of DAM method is described, as follows.

Without a loss of generality, the received IF signal can be represented by

$$S(t) = \sqrt{2P_s}D(t)C(t)\cos\left(2\pi f_c t + \varphi\right) + N(t) \tag{1}$$

$P_s$ is the power of the signal. $D(t)$ represents the navigation message and the Neumann–Hoffman code. $D(t)=\pm1$. The bit sign transition occurs when $D(t)$ changes from 1 to -1 and vice versa. $C(t)$ represents the ranging code. $C(t)=\pm1$. $f_c$ is the carrier frequency, $f_c=f+\Delta f$. $f$ is the IF frequency. $\Delta f$ is the frequency offset caused by Doppler effect and the local oscillator deviation. Generally, $\Delta f$ is much smaller than $f$. $\varphi$ is the carrier initial phase. $N(t)$ is the additive white Gaussian noise (AWGN). $N(t)$ follows a Gaussian distribution with a mean of 0 and variance of $\sigma^2$. The signal-to-noise ratio (SNR) of the received signal can be expressed as

$$SNR = 10\lg\left(\frac{P_s}{\sigma^2}\right) \tag{2}$$

If the IF signal $S(t)$ is delayed by time $\tau$, the result is

$$S(t-\tau) = \sqrt{2P_s}D(t-\tau)C(t-\tau)\cos\left(2\pi f_c(t-\tau)+\varphi\right) + N(t-\tau) \tag{3}$$

By multiplying $S(t)$ with the delayed signal $S(t-\tau)$, we create a new signal, as below.

$$S_\tau(t) = S(t)S(t-\tau) = P_s D_\tau(t)C_\tau(t)\cos(2\pi f_c \tau) + N_\tau(t) \tag{4}$$

$$D_\tau(t)=D(t)D(t-\tau) \tag{5}$$

$$C_\tau(t)=C(t)C(t-\tau) \tag{6}$$

$$\begin{aligned}N_\tau(t)=&-P_s D_\tau(t)C_\tau(t)\cos\left(4\pi f_c(t-\tau/2)+2\varphi\right)\\&+\sqrt{2P_s}\,D(t)C(t)\cos\left(2\pi f_c t+\varphi\right)N(t-\tau)\\&+\sqrt{2P_s}\,D(t-\tau)C(t-\tau)\cos\left(2\pi f_c(t-\tau)+\varphi\right)N(t)\\&+N(t)N(t-\tau)\end{aligned} \qquad (7)$$

As shown in Figure 1, $D_\tau(t)$ is equal to 1 at most of the time except the moment of bit sign transitionn of $S(t)$, as shown in Figure 1. If $\tau \ll T_0$, it can be approximated that $D_\tau(t) \equiv 1$. While considering that $f_c$ is approximately equal to $f$, $\cos(2\pi f_c \tau)$ is approximately equal to $\cos(2\pi f \tau)$, which is a constant. It is possible to choose a favorable delay $\tau$ to maximize $|\cos(2\pi f \tau)|$ to unity. Above all, Equation (4) can be simplified as

$$S_\tau(t) = P_s \cos(2\pi f \tau)C_\tau(t) + N_\tau(t) \qquad (8)$$

This signal does not have any carrier component or bit sign transition. The new signal $S_\tau(t)$ is used for correlating with the new ranging code which is the product of the ranging code $C(t)$ and a $\tau$ delayed version of itself in order to find the beginning of ranging code. Once the beginning of ranging code is detected, it is easy to find the carrier frequency by performing a frequency estimation.

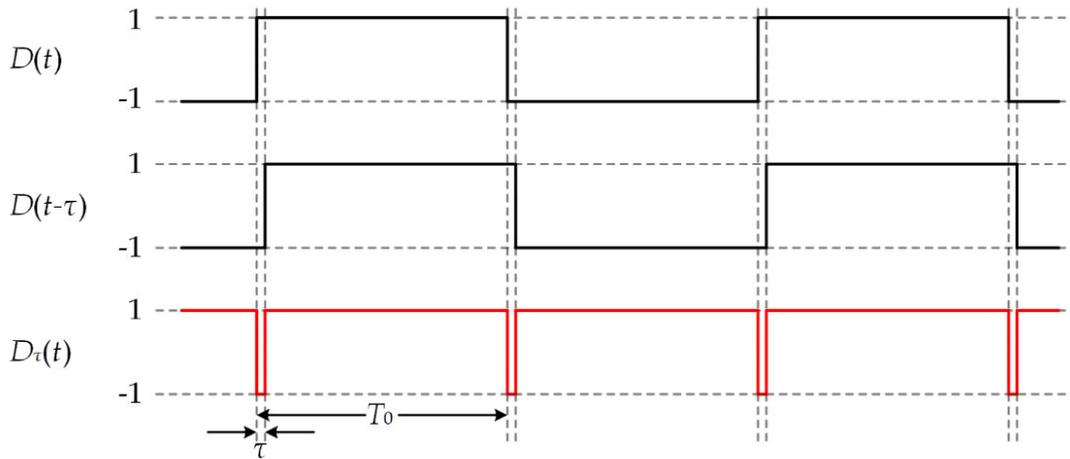

**Figure 1.** The bit sign transitions are eliminated by delay and multiplication (DAM) operation

It is crucial to determine an appropriate value of $\tau$ since the delay $\tau$ is directly related to the SNR of $S_\tau(t)$ and the correlation performance of $C_\tau(t)$. In the following section, we will research the correlation performance of the new ranging code $C_\tau(t)$ and the SNR of the new signal $S_\tau(t)$.

2.1.1. Correlation Performance of the New Ranging Code

In the acquisition of satellite signals, the autocorrelation performance of the ranging code helps to find the beginning of code, and the cross-correlation performance between different ranging codes can prevent the interference of other satellite signals. Therefore, the correlation performance of the ranging codes directly affects the success rate of acquisition.

Suppose that $\{C^i(t)\}$ is a set of ranging codes. Every ranging code is broadcast by different satellites. The autocorrelation function of each ranging code can be expressed as

$$R_i(\Delta t) = \frac{1}{T_0}\int_0^{T_0} C^i(t)C^i(t-\Delta t)\,\mathrm{d}t \qquad (9)$$

where $T_0$ is the period of ranging code. The autocorrelation performance of a ranging code can be measured by the ratio of the peak value ($\Delta t=0$) to the second peak value. The larger the ratio, the better the autocorrelation performance. Thus, the autocorrelation performance of a set of ranging codes can be described as

$$K_{auto} = \min_{i} \frac{R_i(0)}{\max_{\Delta t \neq 0} R_i(\Delta t)} \quad (10)$$

Similarly, the cross-correlation function between every ranging code can be expressed as

$$R_{i,j}(\Delta t) = \frac{1}{T_0} \int_0^{T_0} C^i(t)C^j(t-\Delta t)dt, \ i \neq j \quad (11)$$

Additionally, the cross-correlation performance of a set of ranging codes can be described as

$$K_{cross} = \min_{i,j} \frac{R_i(0)}{\max_{\Delta t} R_{i,j}(\Delta t)} \quad (12)$$

We calculate the $K_{auto,\tau}$ and $K_{cross,\tau}$ of the new ranging codes $\{C_\tau^i(t)\}$ in the BeiDou B1 signal. Figure 2 shows these results. These ratios vary with the increases of $\tau$. Besides, the $K_{auto,ref}$ and $K_{cross,ref}$ of the origin ranging codes $\{C^i(t)\}$ are also shown as references. When $\tau$ is within certain ranges, such as [0.48 µs, 0.55 µs], [0.9 µs, 1.05 µs], [1.4 µs, 1.53 µs], the set of new ranging codes $\{C_\tau^i(t)\}$ even shows a better correlation performance than the origin codes $\{C^i(t)\}$.

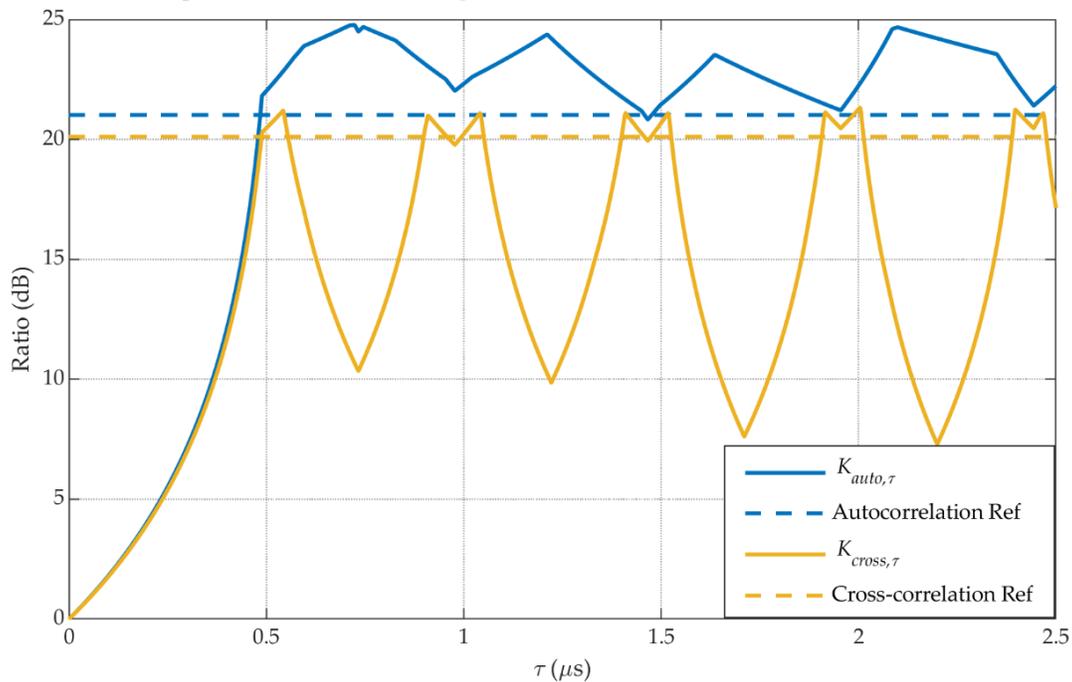

**Figure 2.** The correlation performance of the new ranging codes and the origin codes

2.1.2. Noise Performance

Although the DAM operation can eliminate the bit sign transition and carrier, it will increase the power of noise greatly. This section will research the noise performance of the DAM operation.

The noise consists of three components, among which the last component is the dominant one because the amplitude of the noise is much larger than the amplitude of the signal in $S(t)$, as described in Equation(7). Thus, Equation(7) can be approximated as

$$N_\tau(t) = N(t)N(t-\tau) \tag{13}$$

where $N(t)$ and $N(t-\tau)$ follow a Gaussian distribution with a mean of 0 and a variance of $\sigma^2$. $N(t)$ and $N(t-\tau)$ are band-limited white noises with a center frequency of $f$ and a bandwidth $B$ due to the role of the RF front-end filter. When $N(t)$ and $N(t-\tau)$ are uncorrelated, the power of noise is the smallest [30]. The correlation between $N(t)$ and $N(t-\tau)$ is given by the autocorrelation function $R(\tau)$.

$$R(\tau) = N_0 B \frac{\sin(\pi B \tau)}{\pi B \tau} \cos(2\pi f \tau) \tag{14}$$

For two Gaussian random variables, $N(t)$ and $N(t-\tau)$, irrelevance is equivalent to independence. In other words, $N(t)$ and $N(t-\tau)$ are independent when $R(\tau)=0$. The product of two independent zero-mean Gaussian random variables follows a distribution of probability density function $P(u)$, as is proven in [30].

$$P(u) = \frac{K_0(|u|/\sigma^2)}{\pi \sigma^2}, \quad u = N(t)N(t-\tau) \tag{15}$$

where $K_0$ is the modified Bessel function of second kind with order 0. The mean and variance of the random variable $u$ is 0 and $\sigma^4$, respectively. Thus, the power of $N_\tau(t)$ is

$$P_{N_\tau} = E(u^2) = [E(u)]^2 + D(u) = \sigma^4 \tag{16}$$

If $|\cos(2\pi f \tau)|=1$, the SNR (in dB) of $S_\tau(t)$ is

$$\text{SNR}_0 = 10\lg\left(\frac{P_s^2}{\sigma^4}\right) = 2\text{SNR} \tag{17}$$

Above all, when $R(\tau)=0$ and $|\cos(2\pi f \tau)|=1$, the power of noise $N_\tau(t)$ is minimum and the power of signal $S_\tau(t)$ is maximum. Equation(17) is strong evidence of why the DAM method is only applicable for pretty strong signals. The SNR (in dB) will be doubled after DAM operation. If the signal is not strong enough, the signal after DAM operation will be drowned in mass noise.

To make a rough estimation of $\text{SNR}_0$, the minimum received power levels on the ground of the BeiDou B1 signal are specified to be -163dBW [4,5]. Supposing that the antenna equivalent noise temperature $T_e$ is 290 K, the minimum SNR of received Beidou B1 signal is

$$\text{SNR} = -163 - 10\lg(kT_e B) \approx -25\text{dB} \tag{18}$$

where $k$ is the Boltzmann constant, $k=1.38\times10^{-23}$, $B$ is the bandwidth of the RF front-end, $B=4$ MHz. Therefore, the $\text{SNR}_\tau$ is

$$\text{SNR}_\tau = 2\text{SNR} = -50\text{dB} \tag{19}$$

It will take long-term coherent integration to obtain enough gain to acquire such a weak signal. In general, the baseband SNR shall be more than 14dB for effective acquisition [28].

$$SNR_\tau + 10\lg(BT_c) \geq 14\text{dB} \tag{20}$$

Solve the inequality, the coherent integration length $T_c \geq 628$ ms, which means that the coherent integration should be at least 628ms to ensure successful acquisition. Besides, there are many factors that may further impact the SNR, such as signal blocking, quantization noise, etc. The coherent integration time should be longer if these factors are taken into account.

2.1.3. The Optimal Delay τ

The value of delay τ will bring a huge impact on the correlation performance of $C_\tau(t)$, the signal power, and the noise power, as analyzed in Section 2.1, 2.1.1, and 2.1.2. In addition, in a sampled system, τ must be an integer multiple of the sampling period. The optimal value of τ shall meet these criterions in Equation(21).

$$\begin{cases} \tau \ll T_0 \\ K_{auto}(\tau) > K_{auto,ref} \\ K_{cross}(\tau) > K_{cross,ref} \\ |\cos(2\pi f \tau)| = 1 \\ \sin(\pi B \tau) = 0 \\ \tau = \Delta n / f_s \end{cases} \tag{21}$$

For example, suppose that the sampling frequency $f_s$=10MHz, $f$=2.5MHz, and $B$=4MHz, the optimal value of τ is 1*μs*. Thus, the number of delayed sampling points $\Delta n$ is 10.

2.2. Acquistion Scheme of VLDA

After DAM operation, it takes long-term coherent integration to acquire satellite signals. If directly performing coherent integration with local ranging code, the computational complexity will be too high to fulfil. As shown in Equation(8), the $S_\tau(t)$ consists of the new ranging code $C_\tau(t)$ and noise $N_\tau(t)$. While considering that $C_\tau(t)$ is a periodic signal, the signal intensity can be enhanced by periodic accumulation. However, the period of $C_\tau(t)$ is not a constant. The period of $C_\tau(t)$ is equal to that of the origin ranging code $C(t)$. The period will be slightly shifted due to the Doppler effect resulting by satellites moving at high speed relative to the ground. The maximum Doppler frequency shift of the ranging code can be calculated by the following equation [28].

$$\Delta f_{max} = V_{sat} \frac{R_e}{R_{sat}} \frac{f_R}{c} \tag{22}$$

$V_{sat}$ is the satellite's speed in the Earth-Centered Earth-Fixed (ECEF) coordinate system. $R_e$ and $R_{sat}$ are the radius of the earth and satellite, respectively. $f_R$ is the ranging code rate. $c$ is the speed of light. Table 2 provides the details about the max Doppler shift of ranging codes with different satellites in the BeiDou system. The code chip width will vary with the Doppler shift. The change cannot be ignored in long-term coherent integration.

**Table 2.** The max Doppler shift of the BeiDou B1/B3 signals' ranging codes

| Satellite Type | GEO | IGSO | MEO |
| --- | --- | --- | --- |
| BeiDou B1 signal | 0.09Hz | 2.89Hz | 5.89Hz |
| BeiDou B3 signal | 0.92Hz | 28.9Hz | 58.9Hz |

While considering the M periods ranging code signal, the number of sampling points are

$$N_s = \frac{MLf_s}{f_R + \Delta f} \tag{23}$$

where $L$ is the ranging code length and $f_s$ is sampling rate. When compared with $\Delta f=0$, the change of sampling points is

$$\Delta n = \frac{MLf_s}{f_R + \Delta f} - \frac{MLf_s}{f_R} = -\frac{MLf_s \Delta f}{(f_R + \Delta f)f_R} \tag{24}$$

The number of sampling points will increase or decrease one point every $N_{per}$ points due to the Doppler shift. The $N_{per}$ is

$$N_{per} = \frac{N_s}{|\Delta n|} = \frac{f_R}{|\Delta f|} \tag{25}$$

We can attempt to insert a sampling point every $N_{per}$ samples ($\Delta f<0$), or delete a sample every $N_{per}$ samples ($\Delta f>0$), to compensate the signal length change caused by Doppler shift. Subsequently, the signal can be periodically accumulated to improve the signal strength. This is the proposed method that we called variable length data accumulation (VLDA).

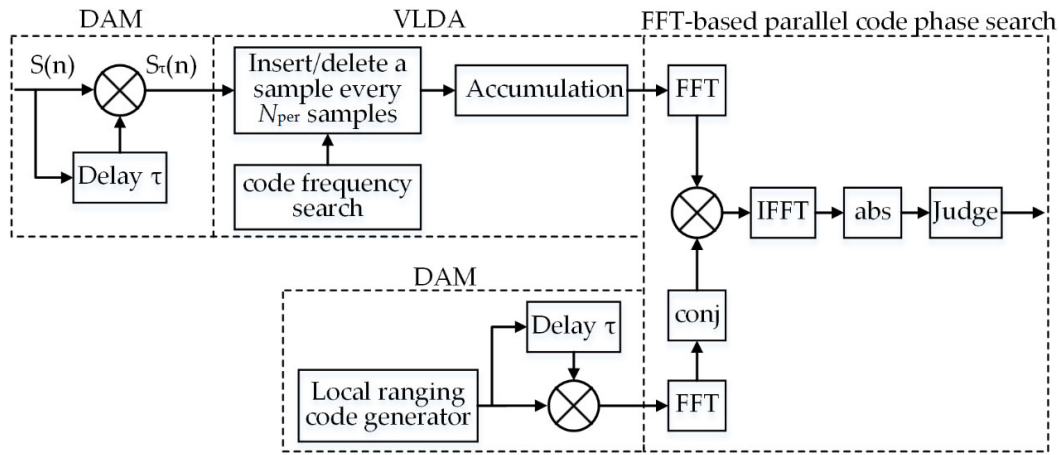

**Figure 3.** The DAM-based variable length data accumulation (VLDA) acquisition scheme

The DAM-based VLDA acquisition scheme consists of three steps, as shown in Figure 3. Firstly, the sampled IF signal $S(n)$ is delayed and multiplied, as well as the local generated ranging code. Secondly, search for the Doppler shift of ranging code. Insert ($\Delta f < 0$) or delete ($\Delta f > 0$) a sample every $N_{per}$ samples, and then periodically accumulated the signal. Finally, the FFT-based parallel code phase search algorithm is adopted to fulfil coherent integration.

2.2.1. The Doppler Search Bins

In the VLDA method, the Doppler shift of the ranging code is searched to find the correct value. The Doppler search range can be set as [-6Hz,6Hz] in the acquisition of BeiDou B1 Signal, according to Table 2. On the other hand, the step of frequency search should be taken into account because the correlation peak is sensitive to frequency error. The simulation result shows that the correlation peaks become narrower as the length of the coherent integration time increases.

The longer coherent integration, the narrower the correlation peaks, as shown in Figure 4. When the coherent length *T*=1s, 2s, 4s, 8s, and 16s, the half-peak widths are 0.72Hz, 0.35Hz, 0.17Hz, 0.08Hz, and 0.04Hz, respectively. In general, the half-peak width is halved when the coherent length is doubled. Therefore, the steps of the Doppler frequency search should be smaller correspondingly. The empirical value of the Doppler frequency search step is

$$\Delta f_{step} = \frac{1}{2T} \tag{26}$$

If the Doppler frequency search range is [$f_{min}$, $f_{max}$], then the number of search bins is

$$N_f = \frac{f_{max} - f_{min}}{\Delta f_{step}} = 2T(f_{max} - f_{min}) \tag{27}$$

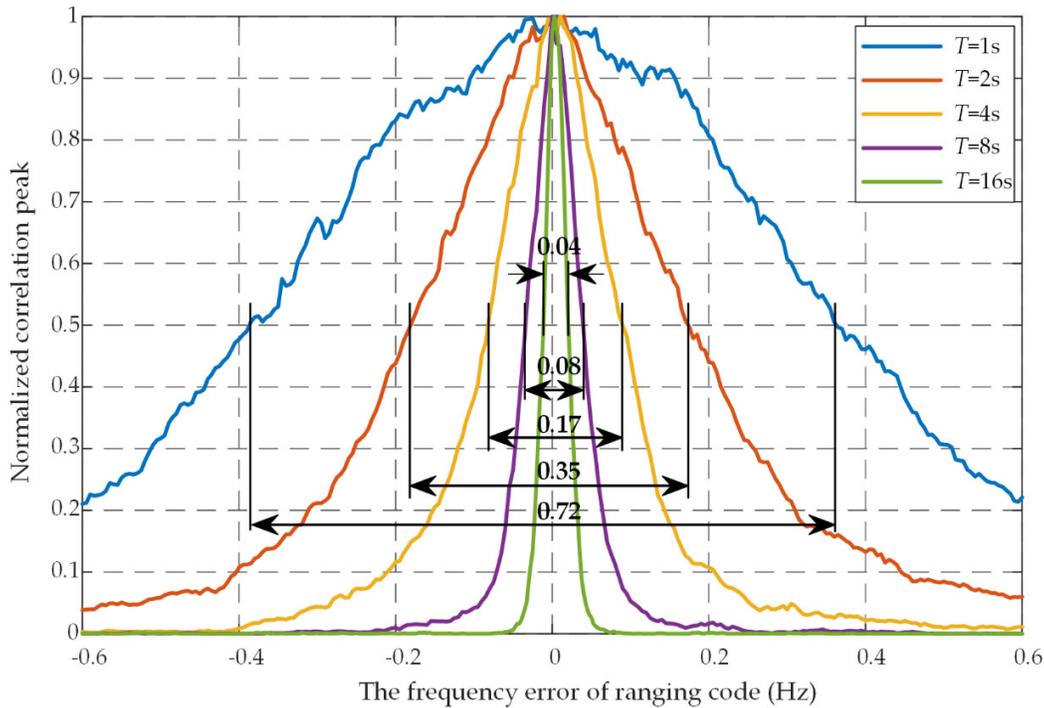

**Figure 4.** The normalized correlation peak versus on the frequency shift of ranging code under different coherent integration time length.

2.2.2. Analysis of the Computational Complexity

In this section, we will analyze the computational complexity of the DAM-based VLDA acquisition scheme. The times of real multiplication operations and real addition operations are considered to be the indicators of the computational complexity. For simplicity, we do not distinguish between integer arithmetic and floating-point arithmetic. Figure 5 is the flow chart of a complete acquisition procedure when using DAM-based VLDA methods.

The left side of Figure 5 shows the times of multiplication operations and addition operations in each step. The red box in the figure represents complex operations. All of the complex operations are converted to real operations. It should be noted that the input data of FFT and the output data of IFFT are real numbers. Therefore, the amount of calculation is reduced by half when compared to complex data. The total times of real multiplication operations and real addition operations are

$$M_{VLDA} = M[N_T + 4N_f N_{sat} + 2N_f N_{sat} \log_2(M)] \tag{28}$$

$$A_{VLDA} = MN_f[N_T + 2N_{sat} + 3N_{sat}\log_2(M)] \tag{29}$$

where $M$ is the number of sampling points for one period ranging code, $M=f_sT_0$, $T_0$=1ms. $N_T$ is the number of accumulated ranging code periods, $N_T=T/T_0$. $N_f$ is the number of frequency search bins. $N_{sat}$ is the number of all satellites in BeiDou system, $N_{sat}$ = 63.

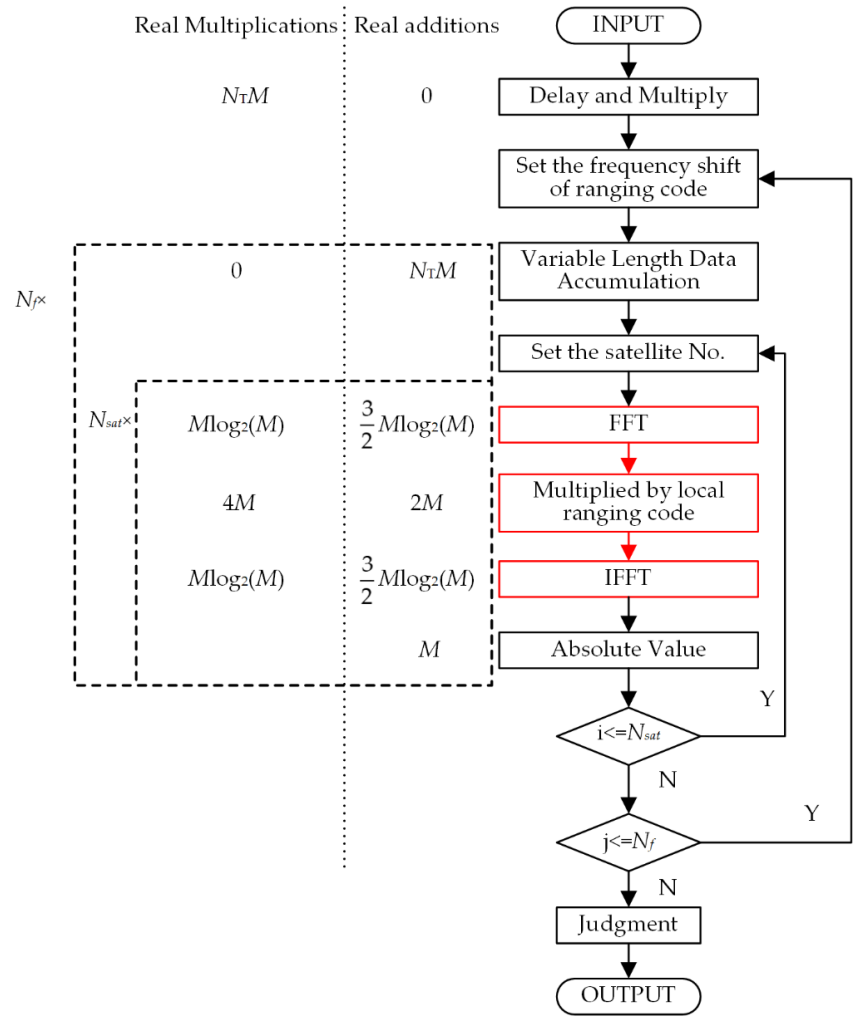

**Figure 5.** The flow diagram and the computational amount of the DAM-VLDA acquisition scheme

As a comparison, Figure 6 shows the flow diagram and computational amount of the NCH acquisition scheme. The times of real multiplication operations and real addition operations are

$$M_{NCH} = MN_fN_{nch}[2 + 6N_{sat} + 4N_{sat}\log_2(M)] \tag{30}$$

$$A_{NCH} = MN_fN_{sat}N_{nch}[4 + 6\log_2(M)] \tag{31}$$

where $N_{nch}$ is the NCH integration length. $N_f$ is the number of carrier frequency search bins. $N_f$ is determined by the search range and search steps of the carrier frequency shift.

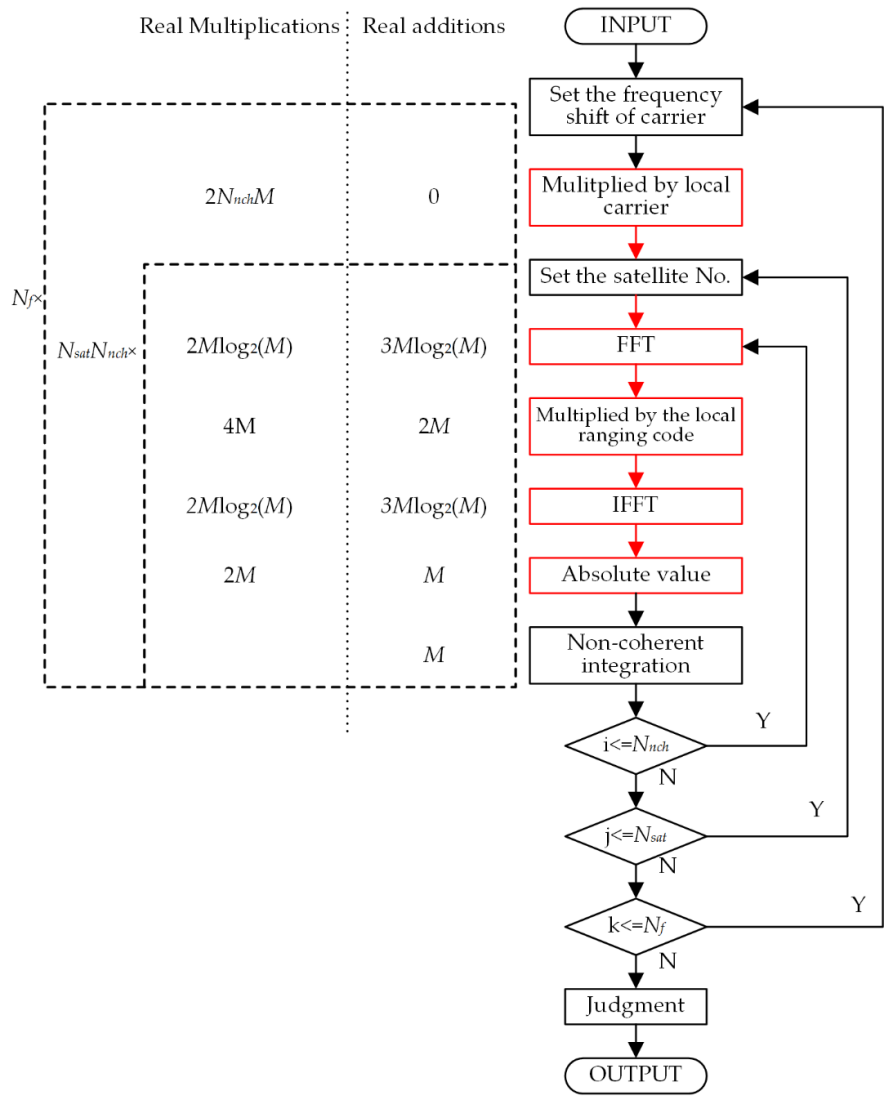

**Figure 6.** The flow diagram and the computational amount of the non-coherent acquisition scheme.

Suppose that the Doppler search range of ranging code is [-6Hz,6Hz] and the carrier frequency shift is [-5kHz,5kHz]; Figure 7 shows the comparison of calculation amount between the VLDA and NCH methods. Four conclusions can be summarized as follows.

1. The calculation amount of two methods increases as the sampling rate increases.
2. The calculation amount of VLDA methods ($T \leq 5$ s) is definitely less than the NCH method ($N_{nch}$=10)
3. It can be approximated that the calculation amount of the VLDA method (T=10 s) is approximately the same as the NCH method ($N_{nch}$=20). For example, the operations of VLDA method (*T*=10 s) are $4.74 \times 10^9$ multiplication and $3.04 \times 10^{10}$ addition if the sampling frequency is 10MHz (black lines in the figures), while those of the NCH method ($N_{nch}$=20) are $1.57 \times 10^{10}$ multiplication and $2.21 \times 10^{10}$ addition. The amount of multiplications is decreased by $1.10 \times 10^{10}$, while that of additions are increased by $8.3 \times 10^9$. The amount of increase and decrease is roughly equal.
4. All in all, the descending order of calculation amount is: VLDA(T=50s) > VLDA(T=20s) > VLDA(T=10s) ≈ NCH(N=20) > NCH(N=10) > VLDA(T=5s) > VLDA(T=2s) > VLDA(T=1s).

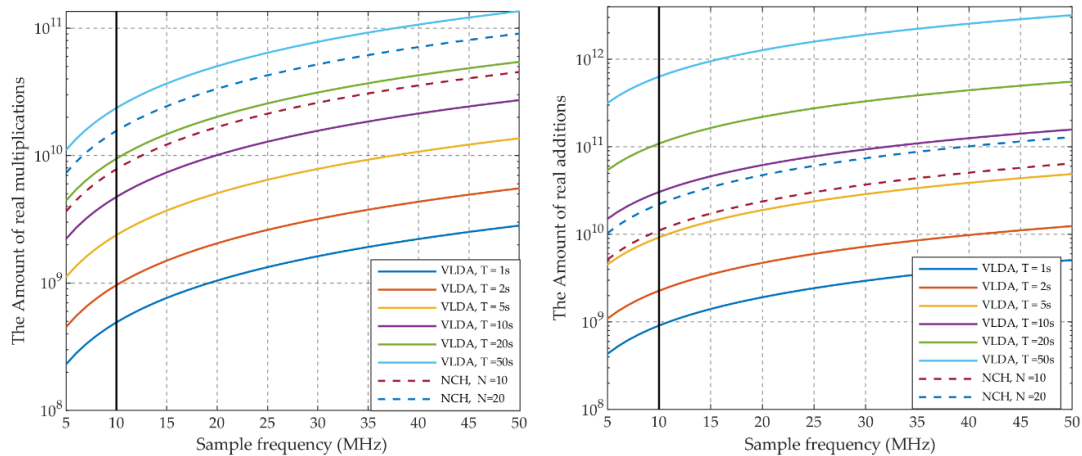

**Figure 7** Comparison of the calculation amount between VLDA and non-coherent (NCH) methods. (**a**) Amount of real multiplications; (**b**) Amount of the real additions.

When compared with the traditional NCH method, the proposed method has low computation complexity. Firstly, the VLDA method performs signal accumulation before FFT/IFFT operations, thus decreasing the number of FFT/IFFT operations. Secondly, in the proposed method, the DAM operation strips off the carrier once and for all. In the traditional NCH method, the carrier in the IF signal is stripped by multiplying local generated carrier, which turns the real signals to become complex signals. Complex signals further increase the computation complexity. Lastly, in the proposed method, DAM and VLDA are all integer arithmetic. The proposed method has lower computation complexity than the NCH method when considering that integer arithmetic is usually faster than float-point arithmetic.

## 3. Results

### 3.1. Simulation Results

Experiments using the Monte-Carlo simulation were carried out to verify the performance of the proposed VLDA scheme as compared with the traditional NCH method on the BeiDou B1 signal receiver. The sampling frequency and intermediate frequency are set as 10MHz and 2.5MHz, respectively. The Doppler frequency search range of carrier and ranging code are set as [-5kHz,5kHz] and [-6Hz,6Hz], respectively. The Doppler frequency shift of carrier and ranging code are set as 1678.6 Hz and 2.2 Hz, respectively. The noises are simulated as AWGN, being randomly generated and added to the simulated BeiDou B1 signals. The results are tested in a computer with Intel Core i7 8700 CPU and 64GB RAM. The simulation programs are high parallel optimized to make full use of the 6-core CPU.

The VLDA methods with different coherent lengths ($T$=1 s, 2 s, 5 s, 10 s, 20 s, 50 s) and the NCH methods with different non-coherent integration numbers ($N$=10, 20) are tested under different $C/N_0$ circumstances, where the interval of $C/N_0$ is 0.25 dB-Hz. Additionally, every simulation was repeated 1000 times to obtain good statistical properties.

Figure 8 shows the probability of detection $P_d$ under the circumstances that the probability of false alarm $P_{fa}$ is $10^{-2}$. The probability of detection refers to the probability that a real satellite signal is successfully acquired. The probability of false alarm refers to the probability that a non-existent satellite signal is mistakenly captured. The proposed VLDA methods has better acquisition sensitivity than the traditional NCH methods with the same amount of computation, as shown in Figure 8. For example, when $C/N_0$=34 dBHz, the $P_d$ of VLDA ($T$=10 s) method is 0.97, while that of the NCH ($N$=20) method is 0.57. With the same amount of calculations, the proposed method improves the probability of detection by 0.4.

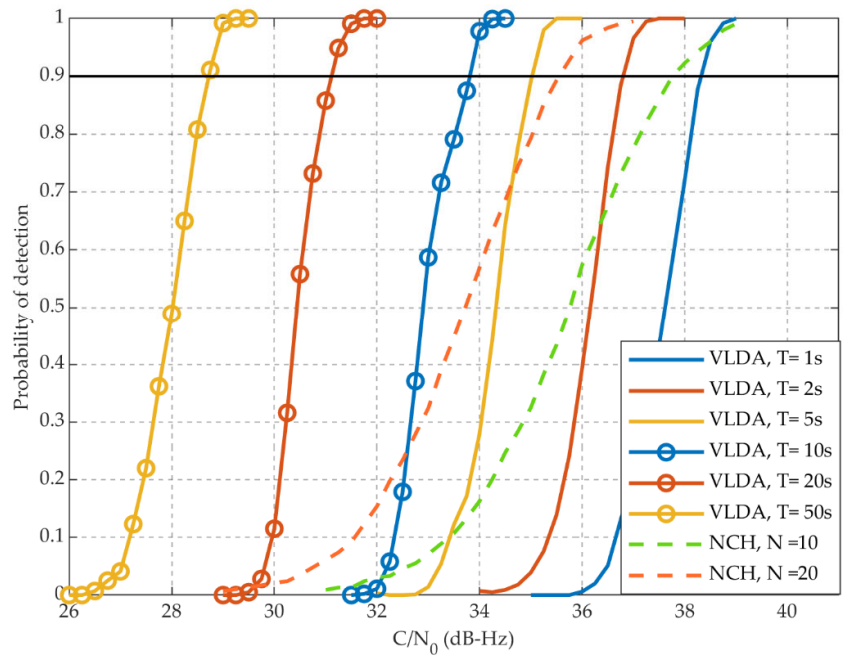

**Figure 8.** The probability of detection $P_d$ under different methods. ($P_f = 10^{-2}$).

Table 3 shows the acquisition sensitivity and computation time of VLDA methods with different coherent integration length. Table 4 shows the acquisition sensitivity and computation time of NCH methods with different non-coherent integration times. The computation time is an average of multiple experiments. The acquisition sensitivity of VLDA (T=5 s) is better than that of NCH (*N*=20), while the computation time can be reduced by 72.5%, as compared with Table 3 and Table 4. Moreover, the acquisition sensitivity of VLDA (*T*=10s) is 1.7dB-Hz higher than that of NCH (*N*=20), while the computation time is almost the same.

**Table 3.** The acquisition sensitivity and computation time of the VLDA methods.

| Coherent integration length | 1s | 2s | 5s | 10s | 20s | 50s |
|---|---|---|---|---|---|---|
| Acquisition sensitivity ($P_d = 0.9$, $P_f = 10^{-2}$) | 38.3dB-Hz | 36.8dB-Hz | 35dB-Hz | 33.8dB-Hz | 31.1dB-Hz | 28.7dB-Hz |
| Computation time | 0.13s | 0.31s | 1.47s | 5.51s | 21.94s | 136.65s |

**Table 4.** The acquisition sensitivity and computation time of NCH methods.

| Non-coherent integration times | 10 | 20 |
|---|---|---|
| Acquisition sensitivity ($P_d = 0.9$, $P_f = 10^{-2}$) | 37.8dB-Hz | 35.5dB-Hz |
| Computation time | 2.67s | 5.35s |

### 3.2. Actual Experimental Verification

An actual experiment was conducted to further verify the proposed VLDA method. A universal software-radio peripheral (USRP) was used as a front-end in BeiDou software receiver, as shown in Figure 9. The USRP can be used for collecting raw IF signal covering the entire family of BeiDou signals [31–32]. In our experiment, the sampling frequency is 10MHz and the intermediate frequency is 2.5MHz. The B1 signal was collected at 37°59.1340'N 117° 20.0979'E. It is located at the College of Electronic Information and Optical Engineering, Nankai University Jinnan Campus, where the antenna was placed on a windowsill in the college building. The Beijing time was 8:39 PM (12:39PM UTC Time), Dec. 31, 2019. Table 5 shows the acquisition result using the VLDA (T=50 s) method. It should be noted that the acquisition only executes once. The number of satellites successfully acquired is 19, including six GEO, nine IGSO, and four MEO.

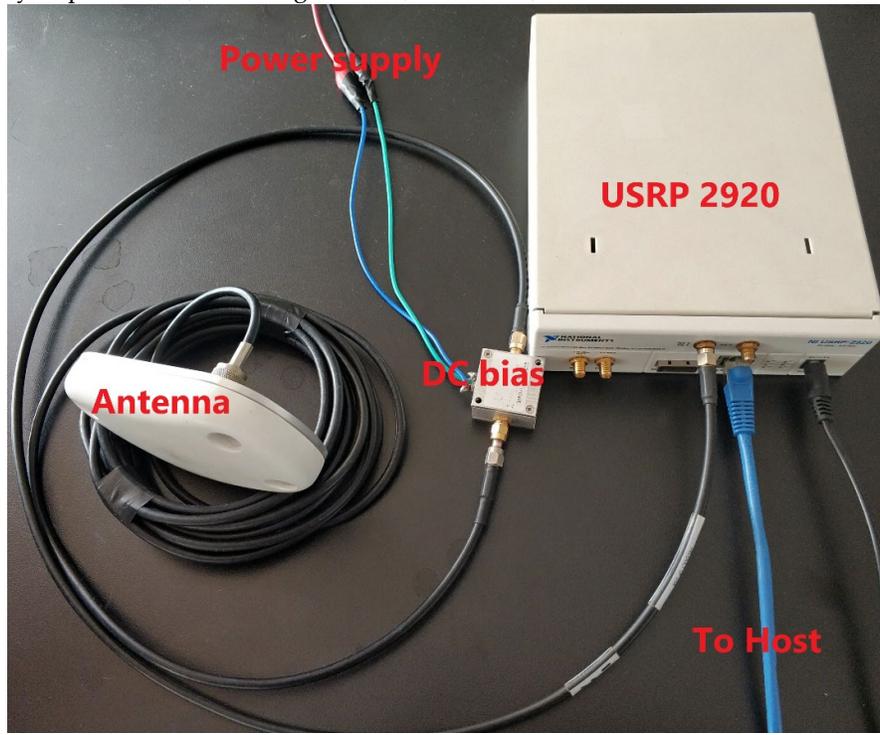

**Figure 9.** The hardware components of the BeiDou software-defined recever front end based on universal software-radio peripheral (USRP).

**Table 5.** The actual acquisition results of BeiDou B1 signal.

| PRN | Peak | Second peak | Ratio | Code phase | Code Frequency (Hz) | Satellite Type |
|---|---|---|---|---|---|---|
| 1 | 256653903 | 27601841 | 9.30 | 3187 | 0.03 | GEO |
| 2 | 76448758 | 32714670 | 2.34 | 2469 | 0.00 | GEO |
| 3 | 177831744 | 32714670 | 5.98 | 2640 | 0.04 | GEO |
| 4 | 105355209 | 32714670 | 3.70 | 6677 | 0.03 | GEO |
| 6 | 55694215 | 32714670 | 3.61 | 2295 | -2.29 | IGSO |
| 7 | 81700227 | 16471847 | 4.96 | 7452 | 0.22 | IGSO |
| 8 | 108572256 | 15557184 | 6.98 | 4391 | 2.40 | IGSO |
| 9 | 50969053 | 13229359 | 3.85 | 7936 | -2.55 | IGSO |
| 10 | 77302355 | 23460561 | 3.29 | 2870 | 0.50 | IGSO |
| 16 | 107220060 | 16684840 | 6.43 | 8341 | -2.42 | IGSO |
| 18 | 107949090 | 27920476 | 3.87 | 8341 | 0.01 | GEO |
| 27 | 623957220 | 44717638 | 13.95 | 4857 | -0.25 | MEO |
| 30 | 623957220 | 28491772 | 14.60 | 6752 | 3.37 | MEO |
| 37 | 682503125 | 44266559 | 15.42 | 7019 | -3.09 | MEO |
| 38 | 682503125 | 23388864 | 10.12 | 2668 | 2.31 | IGSO |
| 39 | 71746708 | 15864686 | 4.52 | 9032 | -2.47 | IGSO |
| 46 | 484922105 | 36872921 | 13.15 | 3657 | 0.47 | MEO |
| 56 | 113757292 | 23939812 | 4.75 | 2667 | 0.54 | IGSO |
| 59 | 288631377 | 28870015 | 10.00 | 4766 | -0.09 | GEO |

The actual experiment proved that the proposed VLDA method was feasible for BeiDou signal acquisition. Although this method is to solve the bit sign transition caused by secondary code, it is also applicable to GEO satellite signals, which are not modulated with secondary code. The VLDA method is a universal approach that can be used for any direct sequence spread spectrum (DSSS) signals.

**4. Discussion**

The simulation results in Section 3.1 prove that the proposed VLDA method has better acquisition sensitivity than traditional NCH method. The VLDA method can improve the sensitivity of 1.7dB under the same calculation amount. In addition, the VLDA method requires only about 27.5% of calculations to achieve the same acquisition sensitivity (35dB-Hz).

The advantages of the proposed VLDA method are summarized, as follows.

- High probability of detection. As long as the $C/N_0$ is greater than a threshold, the VLDA method can successfully acquire signals with 100% probability. As a comparison, the performance of NCH method is related to the initial phase of ranging code and the secondary code, which might lead to acquisition failure, even at high $C/N_0$.
- Low computational complexity. The VLDA method can fulfill the acquisition with a pretty low amount of calculation. This only works for strong signals ($C/N_0>35$dB-Hz).
- Universal applicable to various GNSS signals. The experiment in Section 3.2 validates the effectiveness of the proposed VLDA method for BeiDou GEO/IGSO/MEO satellites. In fact, this method is applicable to various GNSS signals and it can effectively solve the bit sign transition in DSSS systems.

However, the VLDA method also has some weaknesses, which are summarized, as follows.

- Long-time signal required. When compared to the NCH method, which requires a few milliseconds of signal, the VLDA method usually requires several seconds of signal.
- Impractical for very weak signal. Theoretically, the VLDA method can detect any faint signal after a long period of signal accumulation. However, the amount of calculation will increase to an intolerable level. In practice, the VLDA method is not recommended if the $C/N_0$ is less than 28dB-Hz.

**5. Conclusions**

We proposed DAM-based VLDA methods in order to eliminate the bit sign transition that is caused by navigation message and secondary modulated NH codes. The proposed method consists of three steps. Firstly, the BeiDou signals are delayed and multiplied to eliminate the carrier and bit sign transition. Secondly, the long-term signals are periodically accumulated to improve the signal intensity. Finally, the FFT-based parallel code phase search algorithm is used for acquisition. The simulation results indicate that the proposed VLDA method has better acquisition sensitivity than the traditional NCH method under the same calculation amount. The VLDA method only requires about 27.5% of calculations to achieve the same acquisition sensitivity (35 dB-Hz). The actual experimental results verify the feasibility of the VLDA method. It has the advantages of high acquisition success rate, low computational complexity, and wide applicability. It can be applied to the BeiDou B1, GPS L1, and Galileo E1OS signals.

**Author Contributions:** All authors have read and agree to the published version of the manuscript. Conceptualization, H.W. methodology, M.Y.; software, M.Y.; validation, M.Y.; formal analysis, M.Y.; investigation, Q.W. and Y.Z.; resources, Y.Z. and Z.L.; data curation, Q.W.; writing—original draft preparation, M.Y.; writing—review and editing, H.W.; visualization, M.Y; supervision, H.W.; project administration, H.W.; funding acquisition, H.W and Z.L. All authors have read and agreed to the published version of the manuscript.

**Funding:** This research was funded by National Natural Science Foundation of China, grant number 61571244; by Tianjin Research Program of Application Foundation and Advanced Technology, grant number 18YFZCGX00480; by National Natural Science Foundation of China, grant number 61871239.

**Conflicts of Interest:** The authors declare no conflict of interest. The funders had no role in the design of the study; in the collection, analyses, or interpretation of data; in the writing of the manuscript, or in the decision to publish the results.

**References**

1. Yang, Y., Mao, Y., Sun, B. Basic performance and future developments of BeiDou global navigation satellite system. *Satell. Nav.* **2020**, *1*, 1−8
2. China Satellite Navigation Office, China Successfully Launched the 52nd and 53rd BDS Satellites. Available online: http://en.beidou.gov.cn/WHATSNEWS/201912/t20191216_19691.html (accessed on 19 February 2020)
3. Test and Assessment Research Center of China Satellite Navigation Office, Constellation Status. Available online: http://www.csno-tarc.cn/en/system/constellation (accessed on 19 February 2020)
4. China Satellite Navigation Office, BeiDou Navigation Satellite System Signal in Space Interface Control Document Open Service Signal B1I (Version 3.0). Available online: http://en.beidou.gov.cn/SYSTEMS/ICD/201902/P020190227702348791891.pdf (accessed on 16 January 2020)
5. China Satellite Navigation Office, BeiDou Navigation Satellite System Signal in Space Interface Control Document Open Service Signal B3I (Version 1.0). Available online: http://en.beidou.gov.cn/SYSTEMS/ICD/201806/P020180608516798097666.pdf (accessed on 16 January 2020)
6. GPS Directorate Missile Systems Center, NAVSTAR GPS Space Segment/User Segment L5 Interfaces (IS-GPS-705F). Available online: https://www.gps.gov/technical/icwg/IS-GPS-705F.pdf (accessed on 16 January 2020)
7. European Union, European GNSS (Galileo) Open Service Signal-in-Space Interface Control Document Issue 1.3. Available online: https://www.gsc-europa.eu/sites/default/files/sites/all/files/Galileo-OS-SIS-ICD.pdf (accessed on 16 January 2020)
8. Russian Space Systems, JSC. Code Division Multiple Access Open Service Navigation Signal in L3 frequency band (Edition 1.0). Available online: http://russianspacesystems.ru/wp-content/uploads/2016/08/ICD-GLONASS-CDMA-L3.-Edition-1.0-2016.pdf (accessed on 16 January 2020)
9. Bhuiyan, M. Z. H.; Soderholm, S.; Thombre, S.; Ruotsalainen, L.; Kuusniemi, H., Overcoming the Challenges of BeiDou Receiver Implementation. *Sensors* **2014**, *14*, 22082−22098
10. Zhu, C.; Fan, X. N., Weak global navigation satellite system signal acquisition with bit transition. *IET Radar Sonar Nav.* **2015**, *9*, 38−47.
11. Borio, D., M-Sequence and Secondary Code Constraints for GNSS Signal Acquisition. *IEEE T. Aero. Elec. Sys.* **2011**, *47* 928−945.
12. Presti, L. L.; Zhu, X.; Fantino, M.; Mulassano, P. GNSS signal acquisition in the presence of sign transition. *IEEE J. Sel. Top. Signal Process.* **2009**, *3*, 557−570
13. Meng, Q.; Liu, J. Y.; Zeng, Q. H.; Feng, S. J.; Chen, R. Z., Neumann-Hoffman Code Evasion and Stripping Method For BeiDou Software-defined Receiver. *J. Navigation* **2017**, *70*, 101−119
14. Meng, Q.; Liu, J. Y.; Zeng, Q. H.; Feng, S. J.; Xu, R., Efficient BeiDou DBZP-based weak signal acquisition scheme for software-defined receiver. *IET Radar Sonar Nav.* **2018**, *12*, 654−662
15. Smidt J.; Ozafrain S.; Roncagliolo P.; Muravchik C. New technique for weak GNSS signal acquisition. *IEEE Lat Am Trans* **2014**, *12*, 889−894.
16. Lin, D. M; Tsui, J., Comparison of acquisition methods for software GPS receiver. In *Proc. ION GPS 2000*, Salt lake City, UT, USA, 19−22 September 2000, pp. 2385−2390.
17. D. J. R. Van Nee；A. J. R. M. Coenen, New Fast GPS code-acquisition technique using FFT. *Electron. Lett.* 1991, 27, 158−160
18. Tang, P.; Wang, S.; Li, X.; Jiang, Z. A low-complexity algorithm for fast acquisition of weak DSSS signal in high dynamic environment. *GPS Solut.* **2017**, 21, 1427−1441
19. Qiu, J.; Qian, Y.; Zheng, R., Non-coherent, differentially coherent and quasi-coherent integration on GNSS pilot signal acquisition or assisted acquisition. In Proceedings of the 2012 IEEE/ION Position, Location and Navigation Symposium, Myrtle Beach, SC, USA, 23−26 April 2012.
20. Jin T.; Lu F.; Liu Y.; Qin H.; Luo X. Double differentially coherent pseudorandom noise code acquisition method for code-division multiple access system. *IET Signal Process.* **2013**, *7*, 587−597
21. Guo, Y. C.; Huan, H.; Tao, R.; Wang, Y., Long-term integration based on two-stage differential acquisition for weak direct sequence spread spectrum signal. *IET Commun.* **2017**, *11*, 878−886

22. Lin, J. Differentially coherent PN code acquisition with full-period correlation in chip-synchronous DS/SS receivers. *IEEE T. Commun.* **2002**, *50*, 698−702
23. Jin, T.; Yang, J.; Huang, Z.; Qin, H.; Xue, R., Multi-correlation strategies fusion acquisition method for high data rate global navigation satellite system signals. *IET Signal Process.* **2015,** *9*, 623−630
24. Li, H.; Lu, M.; Feng, Z. Partial-correlation-result reconstruction technique for weak global navigation satellite system long pseudo-noise-code acquisition. *IET Radar Sonar Nav.* **2011**, *5*, 731−740
25. Wu, F.; Fu, Y.; Jan, T.; Liang, Y. Weak BeiDou signal acquisition in the presence of sign transition over multiple periods. *Int. J. Satell. Comm. N.* **2018**, *36*, 542–552
26. Ziedan, N. I.; Garrison, J. L., Unaided acquisition of weak GPS signals using circular correlation or double-block zero padding. In Proceedings of the PLANS 2004. Position Location and Navigation Symposium (IEEE Cat. No.04CH37556), Monterey, CA, USA, 26–29 April 2004.
27. Zhang, W.; Ghogho, M., Computational Efficiency Improvement for Unaided Weak GPS Signal Acquisition. *J. Navigation* **2012,** *65*, 363–375
28. Lin, D.; Tsui, J., September. Acquisition schemes for software GPS receiver. Available online: gps-ttff.tripod.com/Schemes.pdf (accessed on 16 January 2020).
29. Tsui, J. B. Y. *Fundamentals of global positioning system receivers: a software approach*, 2nd ed.; Wiley Series in Microwave and Optical Engineering; John Wiley & Sons, Inc.: Hoboken, NJ, USA, 2004.
30. Sharma, R. K.; Wallace, J. W., Improved Spectrum Sensing by Utilizing Signal Autocorrelation. In Proceedings of the VTC Spring 2009 - IEEE 69th Vehicular Technology Conference, Barcelona, Spain, 26–29 April 2009.
31. Peng, S.; Morton, Y. A USRP2-based reconfigurable multi-constellation multi-frequency GNSS software receiver front end. *GPS Solut.* **2013,** *17*, 89–102
32. Peng, S.; Morton, Y. A USRP2-Based multi-constellation and multi-frequency GNSS software receiver for ionosphere scintillation studies. In Proceedings of the 2011 International Technical Meeting of The Institute of Navigation, San Diego, CA, USA, 24–26 January 2011.